\PassOptionsToPackage{table}{xcolor}
\documentclass[acmtog,noacm=true,authorversion=true]{acmart}
\usepackage{svg}
\usepackage{subcaption}
\usepackage{xcolor}  
\usepackage{colortbl}
\usepackage{float}
\usepackage{pifont}
\newcommand{\xmark}{\ding{55}}
\definecolor{yescolor}{HTML}{C6EFCE}
\definecolor{nocolor}{HTML}{FFC7CE}

\newcommand{\cellcoloryesno}[1]{
    \ifnum\pdfstrcmp{#1}{Yes}=0
        \cellcolor{yescolor}{#1}
    \else
        \ifnum\pdfstrcmp{#1}{No}=0
            \cellcolor{nocolor}{#1}
        \else
            {#1}
        \fi
    \fi
}

\renewcommand{\arraystretch}{2.5}

\acmSubmissionID{802}

\AtBeginDocument{%
  }

\acmISBN{978-1-4503-XXXX-X/18/06}

\include{custom_commands}

\begin{document}

\title{Streaming of rendered content with adaptive frame rate and resolution}

\author{Yaru Liu}
\authornote{Equal contribution}
\email{yl962@cam.ac.uk}
\affiliation{%
  \institution{University of Cambridge}
  \city{Cambridge}
  \country{UK}
}

\author{Joseph G. March}
\authornotemark[1] 
\email{joemarch010@gmail.com}
\affiliation{%
  \institution{University of Cambridge}
  \city{Cambridge}
  \country{UK}
}




\author{Rafa{\l} K. Mantiuk}
\email{rafal.mantiuk@cl.cam.ac.uk}
\affiliation{%
  \institution{University of Cambridge}
  \city{Cambridge}
  \country{UK}
}

\copyrightyear{2026}
\acmYear{2026}
\setcopyright{cc}
\setcctype{by}
\acmConference[SIGGRAPH Conference Papers '26]{Special Interest Group on Computer Graphics and Interactive Techniques Conference Conference Papers}{July 19--23, 2026}{Los Angeles, CA, USA}
\acmBooktitle{Special Interest Group on Computer Graphics and Interactive Techniques Conference Conference Papers (SIGGRAPH Conference Papers '26), July 19--23, 2026, Los Angeles, CA, USA}
\acmDOI{10.1145/3799902.3811136}
\acmISBN{979-8-4007-2554-8/2026/07}

\renewcommand{\shortauthors}{Liu et al.}

\begin{abstract}
Streaming rendered content is an attractive way to bring high-quality graphics to billions of mobile devices that do not have sufficient rendering power. Existing solutions render content on a server at a fixed frame rate, typically 30 or 60 frames per second, and reduce resolution when bandwidth is restricted. 
However, this strategy leads to suboptimal rendering quality under the bandwidth constraints. 
In this work, we exploit the spatio-temporal limits of the human visual system to improve perceived quality while reducing rendering costs by adaptively adjusting both frame rate and resolution based on scene content and motion. Our approach is codec-agnostic and requires only minimal modifications to existing rendering infrastructure. We propose a system in which a lightweight neural network predicts the optimal combination of frame rate and resolution for a given transmission bandwidth, content, and motion velocity. This prediction significantly enhances perceptual quality while minimizing computational cost under bandwidth constraints. The network is trained on a large dataset of rendered content labeled with a perceptual video quality metric. The dataset and further information can be found at the project \href{https://www.cl.cam.ac.uk/research/rainbow/projects/adaptive_streaming/}{web page}.


\end{abstract}

\begin{teaserfigure}
  \includegraphics[width=\textwidth]{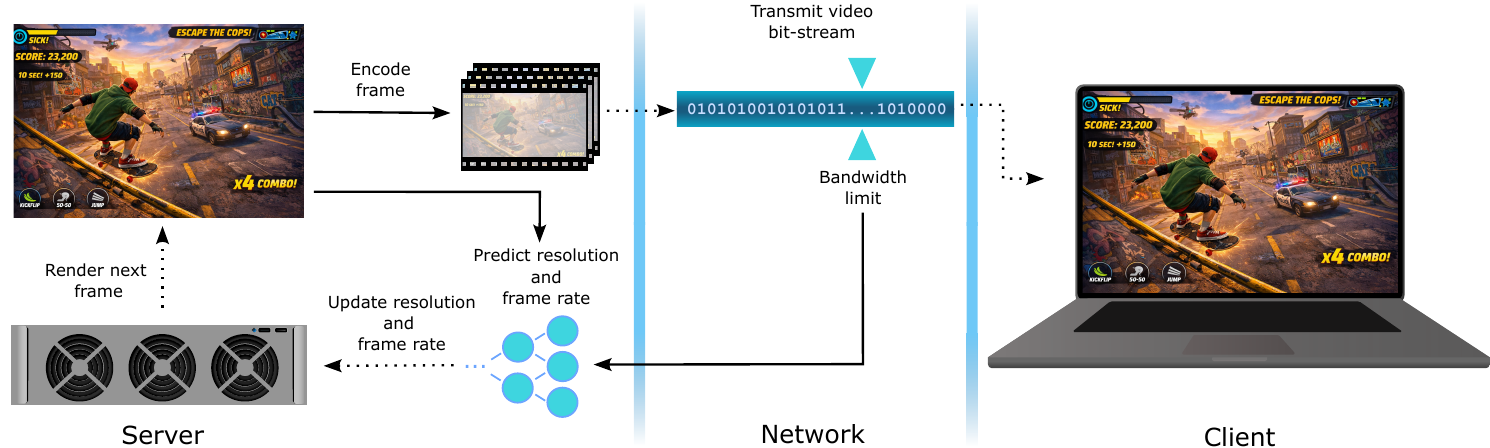}
  \caption{Motivated by the goal of minimizing GPU usage while maintaining high visual quality, we propose a novel real-time method that leverages the human visual system to adaptively adjust the resolution and frame rate of streamed rendered content under bandwidth constraints. By jointly considering image content, motion velocity, and network bandwidth, our method predicts the resolution and frame rate combination that delivers superior perceived quality while reducing rendering costs. 
 }
  \label{fig:teaser}
\end{teaserfigure}


\maketitle

\section{Introduction}
Mobile devices, such as smartphones, tablets and all-in-one AR/VR headsets, have limited graphical compute power due to their energy, thermal and form-factor constraints. An attractive way to circumvent these limitations is to render frames or other relevant data on a server or PC and then stream them to the mobile device. 
Existing game streaming solutions typically rely on available video codecs (h.264/h.265/AV1) and mitigate network bandwidth constraints by selecting the most suitable resolution, while streaming at a fixed frame rate (commonly 60 frames per second). However, this approach results in suboptimal rendering quality under varying bandwidth and motion conditions.

A more effective solution is to adaptively select both the most suitable resolution and frame rate. This is because the fast motion found in real-time rendering can greatly benefit from higher frame rates, even if resolution must be reduced to meet bandwidth limits. Conversely, in low-motion scenarios, it is more efficient to prioritize resolution over frame rate. Fixed-frame-rate solutions cannot make this trade-off and, therefore, offer only suboptimal quality and user experience.
Furthermore, by leveraging the spatio-temporal limits of human vision, we demonstrate that rendering costs can be substantially reduced without impairing perceived quality, making our approach not only perceptually superior but also more computationally efficient.

Adaptive Frameless Rendering (AFR) \cite{dayal_adaptive_2005}, reallocates samples across space and time using temporal reuse and space–time reconstruction to improve quality while minimizing computational effort. However, the approach cannot be easily adapted to modern real-time rendering pipelines. \citeN{denes_perceptual_2020} proposed an adaptive rendering approach in which the frame rate and resolution were selected based on motion velocity and eye-tracking data. \citeN{Jindal_2021} made use of variable rate shading (VRS) to apply motion-adaptive control of the local shading resolution and frame rate. However, both of these works assume that the GPU power is the main bottleneck and do not consider video streaming, instead making the naive assumption that the rendered content is constrained only by the number of pixels rendered per second. 
In game streaming systems, the choice of the best frame rate and resolution depends on the bandwidth of compressed video stream (in bits per second), employed video codec and multiple other variables, which have not been considered before. 

Other works have proposed split-rendering approaches in which one portion of the rendering is performed on the server and another on the client \cite{mueller_shading_2018,Vining_2025_fastatlas}. Split rendering can reduce the required bandwidth and latency. It requires, however, redesigned rendering pipelines, which are often incompatible with the existing game engines and require major infrastructure changes to deploy at scale. 

We present a motion-aware, codec-agnostic extension to existing game streaming systems, which requires no major changes in the rendering pipeline. Our approach uses a neural network that exploits motion velocity, alongside frame content and available bandwidth, to predict the optimal encoding frame rate and resolution. 
A central contribution of this work is an adaptive frame rate and resolution streaming system that leverages the spatio-temporal limits of human vision and a validated video-quality metric to balance GPU rendering cost and perceptual quality under realistic network constraints.

The method yields over 50\% savings in rendering cost, \edit{as measured in pixels rendered per second}, while maintaining high perceived quality. The neural network runs in real time (< 2\,ms) and is trained on a large dataset of rendered video clips. To label the clips, we extended the existing video quality metric, ColorVideoVDP \cite{mantiuk_colorvideovdp_2024}, to handle test and reference videos that differ in frame rate. The dataset will be released prior to publication.

Our main contributions can be summarized as:

\begin{itemize}
\item A method that leverages the spatio-temporal limits of human vision to adaptively select frame rate and resolution, delivering high perceptual quality in streaming while substantially reducing rendering costs.
\item A large \href{https://doi.org/10.17863/CAM.129935}{dataset}\footnote{The dataset: \url{https://doi.org/10.17863/CAM.129935}} of 69,611 rendered with a game engine at different camera velocities, resolutions (from 360p to 1080p), frame rates (from 30\,Hz to 120\,Hz), and encoded at different video compression bandwidths. The dataset lets us determine the optimal frame rate and resolution at a given bandwidth and train a neural network.
\end{itemize}

\vspace{\fill}

\section{Related Work}

\subsection{Content-adaptive video streaming}
\label{sec:2.1} 
Adaptive selection of video encoding parameters is a well-explored area of video streaming. Those methods aim at maximizing visual quality when streaming videos of varying spatio-temporal complexity under constrained or varying network conditions \cite{katsavounidis2018dynamic}.
\citeANP{Bhat2020} \citeNN{Bhat2020} proposed a real-time video encoding resolution predictor. This approach relies on features extracted from the current and previous frame in combination with the resolution selected previously to compute the optimal resolution at which the current video sequence should be encoded.  
Spatio-temporal quality trade-off was considered in the works on the rate-distortion optimization \cite{Vetro_2001} and dynamic frame-rate selection \cite{Thammineni2008}. 

Those works, however, focus on live video streaming, in which frame-rate reduction is achieved by skipping frames. In our work, we assume we can control the frame rate of the renderer, and thus we are able to stream at variable frame rates. Those works also relied on simple empirical formulas that maximized peak-signal-to-noise ratio (PSNR), which is a simple non-perceptual metric. Instead, we employ a perceptual metric (ColorVideoVDP) that models spatio-temporal vision. 

\subsection{No-reference metric of streamed gaming videos}
The rise in popularity of game streaming services, both passive (e.g., Twitch) and active (e.g., GeForce Now), motivated research on the quality of streamed computer graphics content. This field is dominated by non-reference metrics \cite{zadtootaghaj_nr-gvqm_2018,Barman_2019,yu_subjective_2022}, as reference content is difficult to acquire in quantities sufficient for training. Those metrics typically extract a number of features borrowed from existing metrics and train a machine learning model to regress those into subjective quality scores. 

\citeANP{yu_subjective_2022} \citeNN{yu_subjective_2022} collected the LIVE-YouTube Gaming video quality dataset, comprising over 600 user-generated gaming videos and 18\,600 quality ratings. They also evaluated various VQA models on this gaming database. The results indicate that natural and synthetic videos exhibit different statistical distributions, suggesting that non-reference metrics trained on natural images, such as NIQUE, may not perform well on gaming datasets. The superior performance of TLVQM compared to BRISQUE highlights the importance of incorporating motion characteristics when assessing the quality of gaming videos. Furthermore, \citeANP{yu_subjective_2022} highlights that deep-learning-based models can effectively capture the characteristics of synthetic videos, indicating their potential suitability for such applications.

None of the proposed metrics or datasets is suitable for our problem, as they do not model or account for the video frame rate. To address this gap, we generate a large collection of videos with reference at a range of frame rates and resolutions and extend an existing full-reference metric that models spatio-temporal vision to label our dataset.

\begin{figure*}
  \centering
   \includegraphics[width=\linewidth]{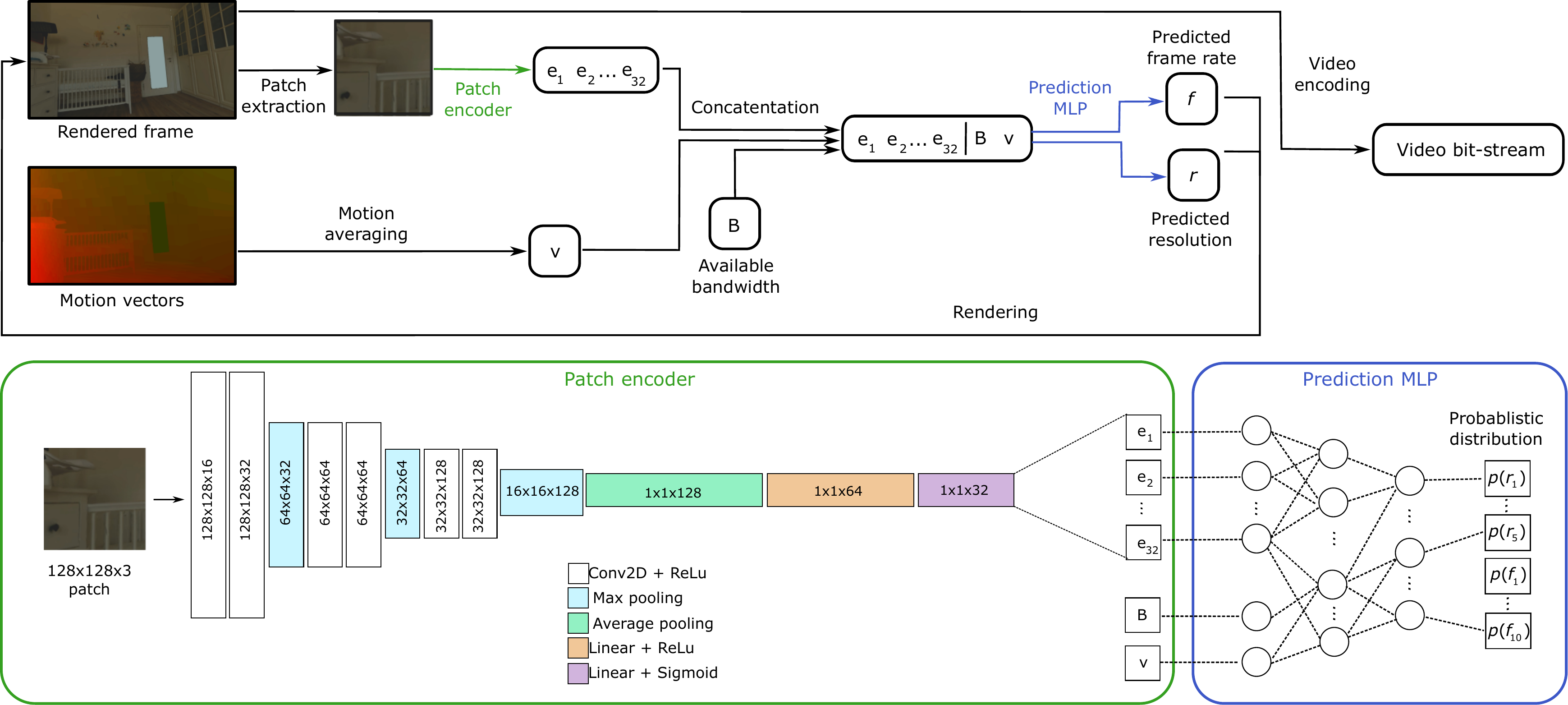}
  \caption{Process diagram for our method. We use standard output from the rendering pipeline: frame data and motion vectors to predict the optimal resolution and frame rate for rendering and video encoding. The bottom of the figure gives architectural overviews of the patch encoder (green) and the MLP used for prediction (blue).}
  \label{fig:process}
\end{figure*}

\subsection{Streaming rendered content}
Previous works designed custom rendering methods for video streaming to attain higher visual quality when streaming rendered content. 

\edit{Streaming texture-space shading data has been explored as an alternative to streaming finished frames \cite{mueller_shading_2018, Vining_2025_fastatlas, Hladky:2021:SnakeBinning, Neff:2022:Meshlets}.} This approach offloads computationally expensive shading to the server while performing visibility calculations on the client. Recent work \cite{Vining_2025_fastatlas} adapted this approach to work within a fixed transmission bandwidth budget while still maintaining high image quality. While effective, such approaches require extensive modifications to typical forward-rendering pipelines. They also require all geometry data to be transferred to and then processed on the client.

\edit{Other works have demonstrated a hybrid approach of streaming limited geometry and image data \cite{hladky_quadstream_2022, Lu:2025:QUASAR}. As with texture-space shading streaming, these approaches render the final image on the client device in a computationally inexpensive manner.}

\citeANP{Liu2015} \citeNN{Liu2015} proposed a method, in which the content's depth buffer were used to compute a saliency map. The saliency map was then used to adjust the quantization parameter of each macroblock in order to redistribute the bandwidth and improve visual quality. We do not consider saliency (which can be unreliable for rendered content) and instead exploit the limitation of spatio-temporal vision to improve visual quality.

\section{Streaming with adaptive frame rate and resolution}
Below, we explain our approach to streaming rendered content with adaptive frame rate and resolution. Because of our focus on real-time rendering, we cannot rely on the techniques used in video streaming, such as a pre-computed resolution/bit-rate ladder \cite{katsavounidis2018dynamic}. 
Instead, we train a real-time predictor that, given video streaming bandwidth, motion velocity, and rendered frames, can predict the optimal combination of frame rate and resolution, maximizing quality while keeping the computational cost and bandwidth limited. 

\begin{figure*}
  \centering
   \includegraphics[width=\linewidth]{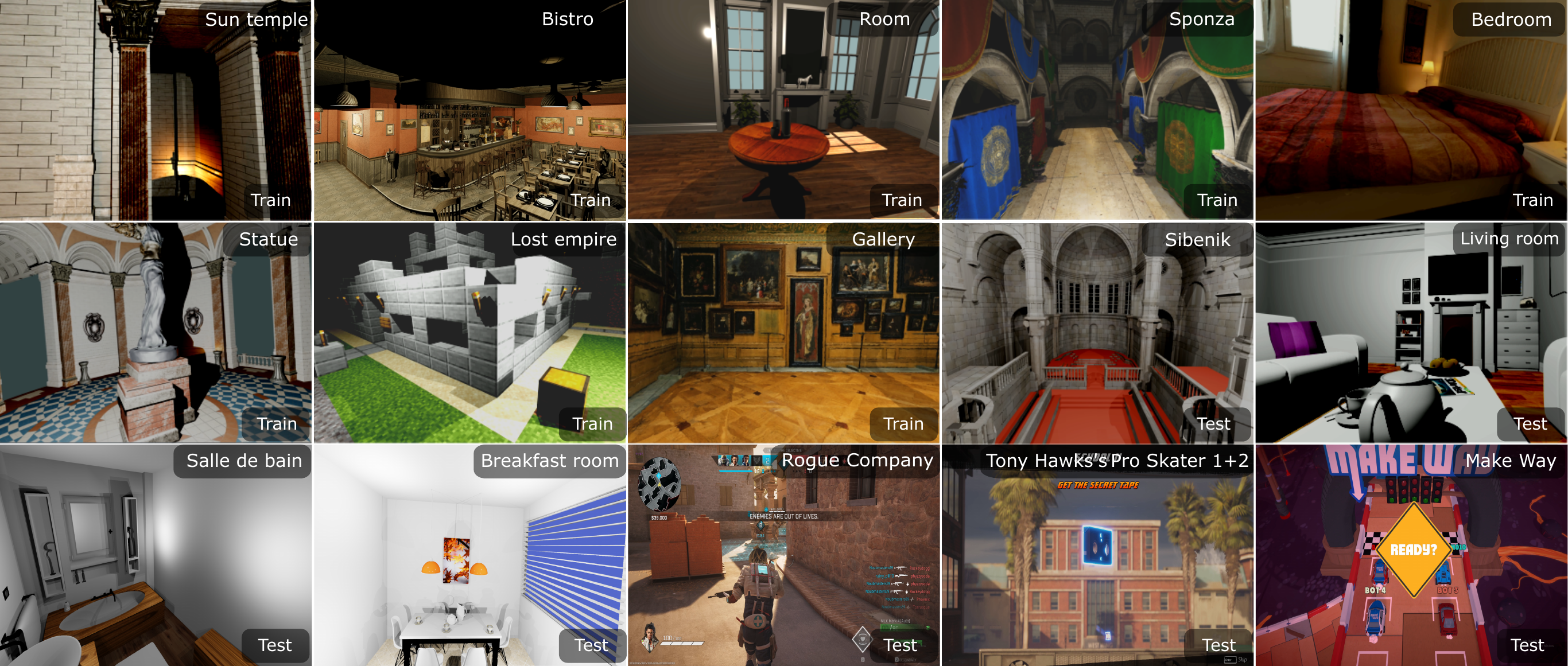}
  \caption{Example frames from the reference videos used for training, testing, and experiments are shown. Sun temple and Statue \cite{OrcaUE4SunTemple} scenes were obtained from Nvidia’s Open Research Content Archive \cite{nvidia_orca_nodate}. Bistro ({\textcopyright}\,2017 Amazon Lumberyard), Room ({\textcopyright}\,Wig42), Sponza ({\textcopyright}\,2010 Frank Meinl, Crytek), Bedroom ({\textcopyright}\,2017 fhernand), Lost empire ({\textcopyright}\,2011 Morgan McGuire), Gallery ({\textcopyright}\,2017 The Hallwyl Museum), Living room ({\textcopyright}\,2012 Jay), Salle de bain ({\textcopyright}\,Nacimus Ait Cherif) and Breakfast room ({\textcopyright}\,Wig42) were downloaded from the McGuire Computer Graphics Archive \cite{noauthor_mcguire_nodate}, and contain only camera motion. Rogue Company (from Rogue Company by First Watch Games), Make Way (from Make Way by Ice BEAM), and Tony Hawks's Pro Skater 1+2 (by 
Vicarious Visions, Iron Galaxy Studios) were captured in-house from gameplay and include both camera motion and dynamic objects. Six scenes were used for testing and experiments; the simplest scene, Living Room, was excluded to keep the experiment duration manageable.}
  \label{fig:scenes}
\end{figure*}

The overview of our adaptive streamed rendering system is shown in \figref{process}. We modified Nvidia's Falcor game engine \cite{nvidia_falcor_2025} to encode, decode, and display the rendered frames. We randomly select a single 128$\times$128 patch from the rendered frame and compute a moving average of frame velocity (last 500\,ms) using motion vectors obtained from the G-buffer. Then, we input the patch, velocity, and current bandwidth into the neural network to predict the resolution–frame rate combination that maintains high perceptual quality while minimizing rendering cost subject to the transmission bandwidth limitation.

\subsection{Real-time rendering system}

To generate the dataset and test our method, we built a proof-of-concept rendering system that integrates the Falcor game engine \cite{nvidia_falcor_2025} with real-time HEVC video encoding and decoding using NVIDIA Video Codec SDK\cite{nvidia_video_sdk}. We render scene content to an 8-bit per channel texture and encode the result into a video stream. We then decode the frame and copy it frame-buffer, up-scaling where necessary using a bilinear filter. Both resolution and frame-rate changes are handled by reconfiguring the state of the encoder and decoder to reflect the desired parameters. Changes to resolution also force the insertion of an I-frame (intra-coded frame, without dependency on neighboring frames). To avoid stuttering during resolution changes, all necessary memory is pre-allocated and shaders are pre-compiled for each potential resolution during renderer initialization. The video decoder output resides in GPU memory, and final color conversion is handled using CUDA to obtain an image in the sRGB color space.

As our predictor only considers network bandwidth, which can be simulated by restricting encoder bandwidth, we do not implement network transmission into our proof-of-concept system. 

\subsection{Adaptive frame rate and resolution dataset}
\label{sec:dataset}

Training our predictor requires a large dataset. We create a dataset of representative video clips that cover a wide variety of rendered content, motion velocities, resolutions, and refresh rates. Then, we need to find the combinations of refresh rate and resolution that deliver the best quality for each clip and use them as labels for our predictor. 

We use 15 scenes, shown in \figref{scenes}, to render 3-second-long video clips. The clips are created by moving a camera along one of 15 fixed paths per scene. Twelve scenes were rendered using Falcor with static objects and moving cameras, while the remaining three were screen-recorded and featured dynamic objects. In total, thirteen scenes—ten rendered and three recorded—were used to train the predictor (see \secref{predictor}), while the final two rendered scenes were reserved exclusively for user studies. We used OBS (version 32.0 \cite{OBS_2026} to record the screen when capturing footage from commercial games. The videos were encoded using the HEVC (H.265) codec, as implemented in libx265, at a constant rate factor (CRF) of 5.

For the rendered scenes used for predictor training and testing, each path is rendered at 3 motion velocities. This gives us 

$10 \times 15 \times 3 = 450$ unique camera clips. Each camera path is rendered at every resolution ($r \in$ \{360, 480, 720, 864, 1080\} lines/height, 16:9 aspect ratio) and frame rate ($f \in$ \{30, 40, … , 120\} Hz) combination and then encoded under one of three bit rates ($b \in$ \{2, 3, 4\} Mbps) with Nvidia's NVENC real-time codec. 
The three recorded scenes have 11 unique camera clips, with slow, medium, and fast velocities. The clips are encoded into 5 resolutions, 10 frame rates, and 3 bit rates as above.
While it is true that encoder bitrate settings may not always be followed exactly by the compression algorithm, in our dataset, the deviation between the target and actual bitrate is negligible. Specifically, the expected error is only 0.63\%, indicating that the bandwidth restrictions were effectively respected.
Encoding is performed with the HEVC (H.265) codec using constant bit rate (CBR) rate control mode and the Main profile. We also generate a reference video for each camera path which is rendered at the highest resolution (1080p), 166\,Hz and encoded using an NVENC encoder at a constant CRF value of 5. This gives us a total of 69,611 video clips.

We selected 1080p as the reference resolution because it remains the standard for streaming across most platforms. While 4K is supported, it is typically reserved for top-tier services and less frequently used due to bandwidth constraints. Moreover, generating and processing the 1080p dataset was computationally expensive, taking approximately a week on a GPU cluster, which made 1080p a practical upper bound for our experiments.
We use 166\,Hz so that none of the test frame rates can be produced by subsampling (in time) by an integer factor. Otherwise, this portion of the frames in the test sequences would coincide in time and be (almost) identical to those in the reference, giving those frame rates an unfair advantage. 

\begin{figure}
   \includegraphics[width=1\linewidth]{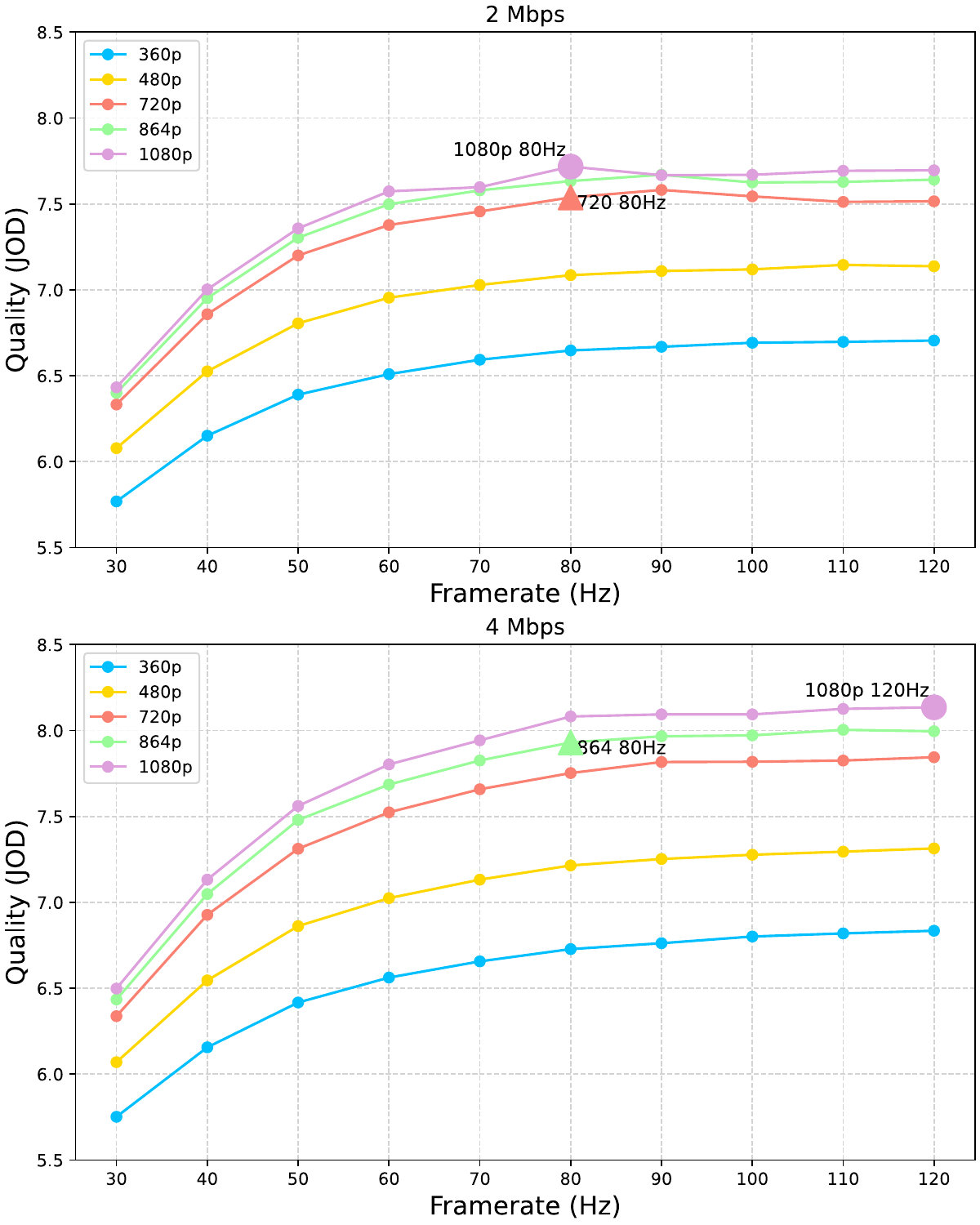}
\caption{ColorVideoVDP predictions for the same sequence from the scene ``Bistro'' rendered at different resolutions and frame rates. The two plots show the results for different bit rates. The resolution and frame giving the highest quality are shown as round markers, and the ones that also reduce the compute cost (computed from \eqref{optimal-f-r}) are shown as triangle markers.}
\label{fig:cvvdp_example}
\end{figure}

\begin{figure}
   \includegraphics[width=1\columnwidth]{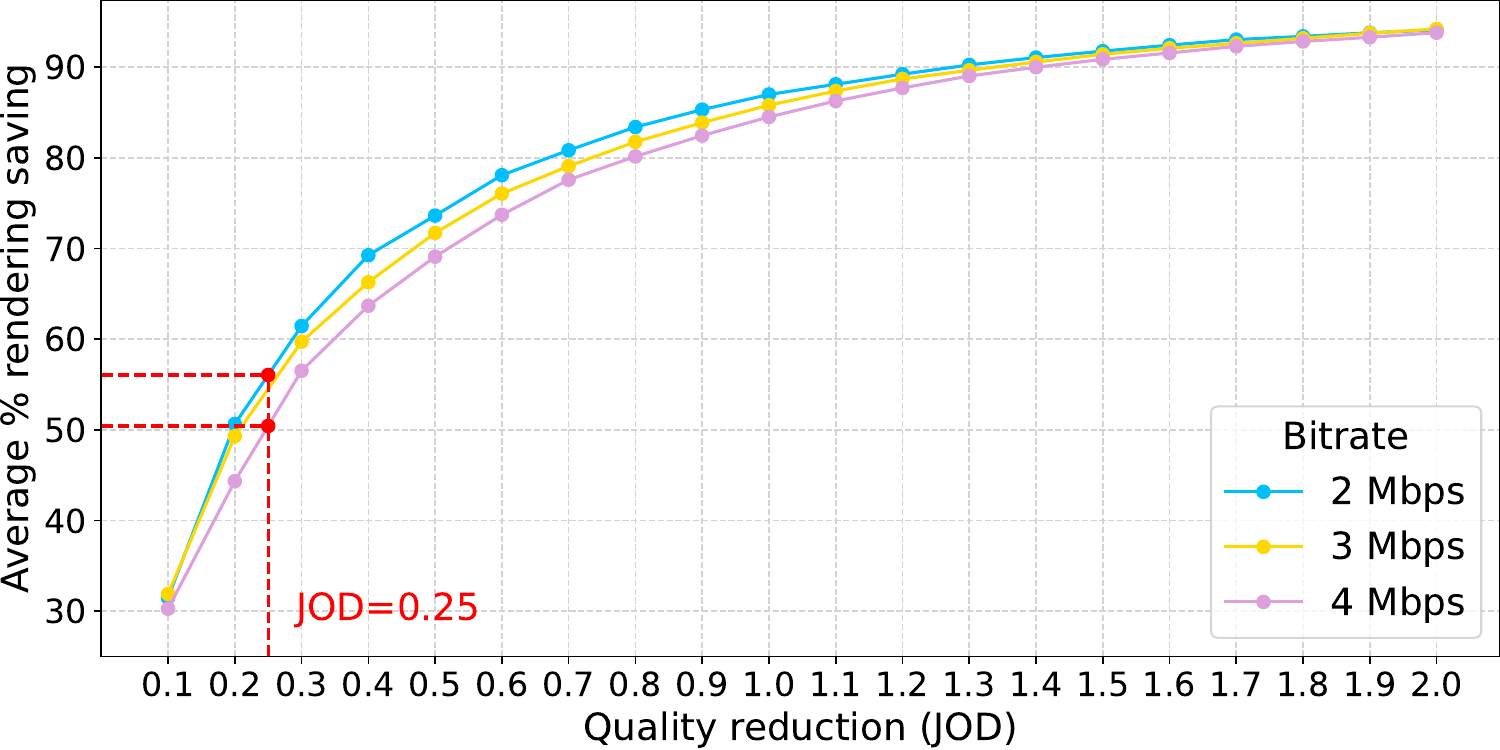}
\caption{Average percentage reduction in pixels rendered per second as we allow for the quality (in JODs) to drop with respect to the maximum quality (see \eqref{optimal-f-r}). Each curve corresponds to one of the considered bit rates.}
\label{fig:dropjod_saving}
\end{figure}

Motion velocity was derived directly from the G-buffer in normalized device coordinates (NDC) by averaging the magnitudes of motion vectors across the frame.
If access to the G-buffer is not possible, such motion vectors could alternatively be obtained with optical flow or by extracting them from MPEG motion vectors. 
It is a common practice to use videos encoded at very high bitrates (or low CRFs) as reference sequences \cite{hammou_effect_2024, chen_hdrsdr-vqa_2025}. The selected CRF is, in practice, visually lossless.

\subsubsection*{Quality labels}
The number of video clips is much too large to measure their quality in a subjective experiment. 
Instead, we extended a recently proposed metric modeling low-level spatio-temporal human vision, ColorVideoVDP \cite{mantiuk_colorvideovdp_2024}, to handle test and reference videos encoded at different frame rates. Our extension of ColorVideoVDP resamples the test videos to match the refresh rate of the reference video --- 166\,Hz. This is achieved by replicating frames between the timestamps of the original test video. Note that the replication of frames accurately simulates video that has a lower frame rate.

An example of ColorVideoVDP predictions for one scene, one path, and one velocity is shown in \figref{cvvdp_example} for two different bit rate settings. The plots show that, depending on the available bandwidth, the highest quality can be obtained when streaming 1080 lines at 80\,Hz or 1080 lines at 120\,Hz (large filled circles in \figref{cvvdp_example}). However, we can also observe that the quality plots tend to flatten near the maximum, suggesting that selecting a slightly different resolution or refresh rate should not affect quality much. Therefore, to minimize the computation load on the server, we selected the combination of the refresh rate and resolution, which resulted in the fewest rendered pixels and was within 0.25 Just-Objectionable-Difference (JOD) units of the maximum: 
\begin{equation}
    f^\ast, r^\ast = \argmin_{f, r} f\,r^2, \quad \text{s.t.} \quad Q^\ast - Q(f,r) \leq 0.25
    \label{eq:optimal-f-r}
\end{equation}

where $f$ is the frame rate, $r$ is the resolution, $Q(f,r)$ is the corresponding JOD quality and $Q^\ast$ is the maximum quality. The choice of 0.25 JOD is validated in Section \ref{sec:exp-jod-reduction}. Large triangles in \figref{cvvdp_example} show the resolution and refresh rate selected using this criterion.
Our validation experiment (\secref{exp-jod-reduction}) will demonstrate that this choice does not affect the quality of streamed content. \figref{dropjod_saving} shows that allowing the quality to be reduced by just 0.25\,JOD allows us to render on average 53\% fewer pixels per second. \edit{We use pixels per second as a naive proxy for GPU utilization as GPU workloads are often shading bound; while this does not capture GPU compute workloads (which can be significant), it is sufficient for our purposes.}

\begin{figure}
  \centering
      \includegraphics[width=\columnwidth]{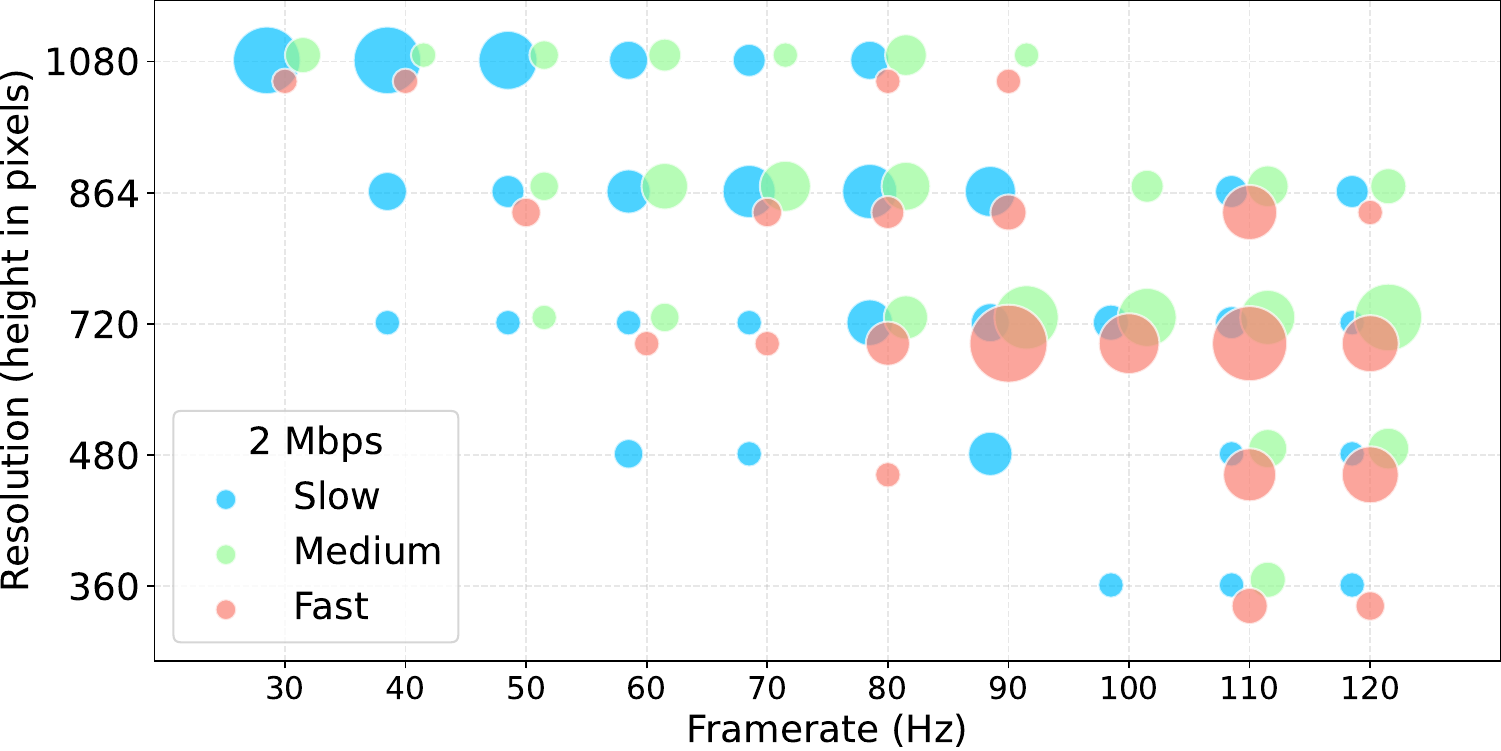}
     \includegraphics[width=\columnwidth]{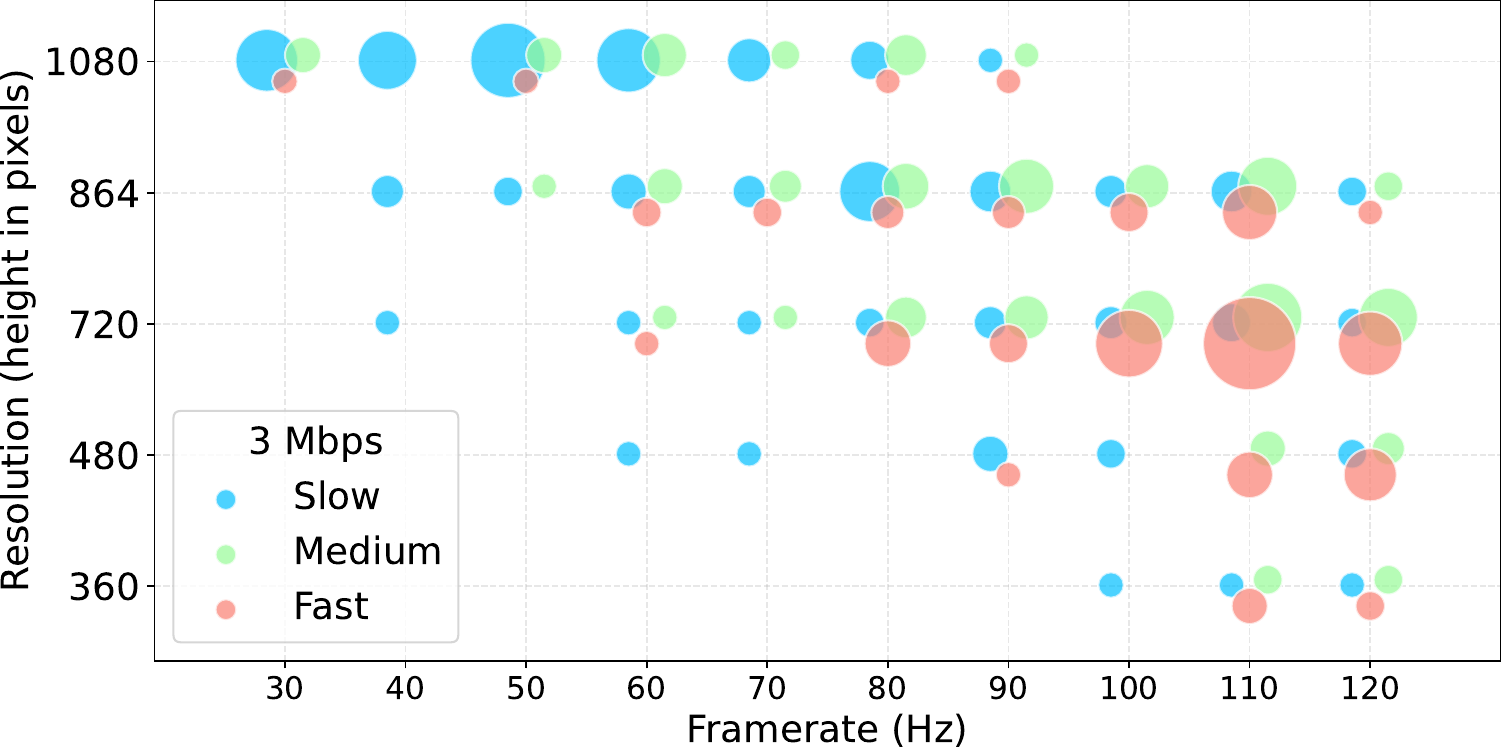}
    \includegraphics[width=\columnwidth]{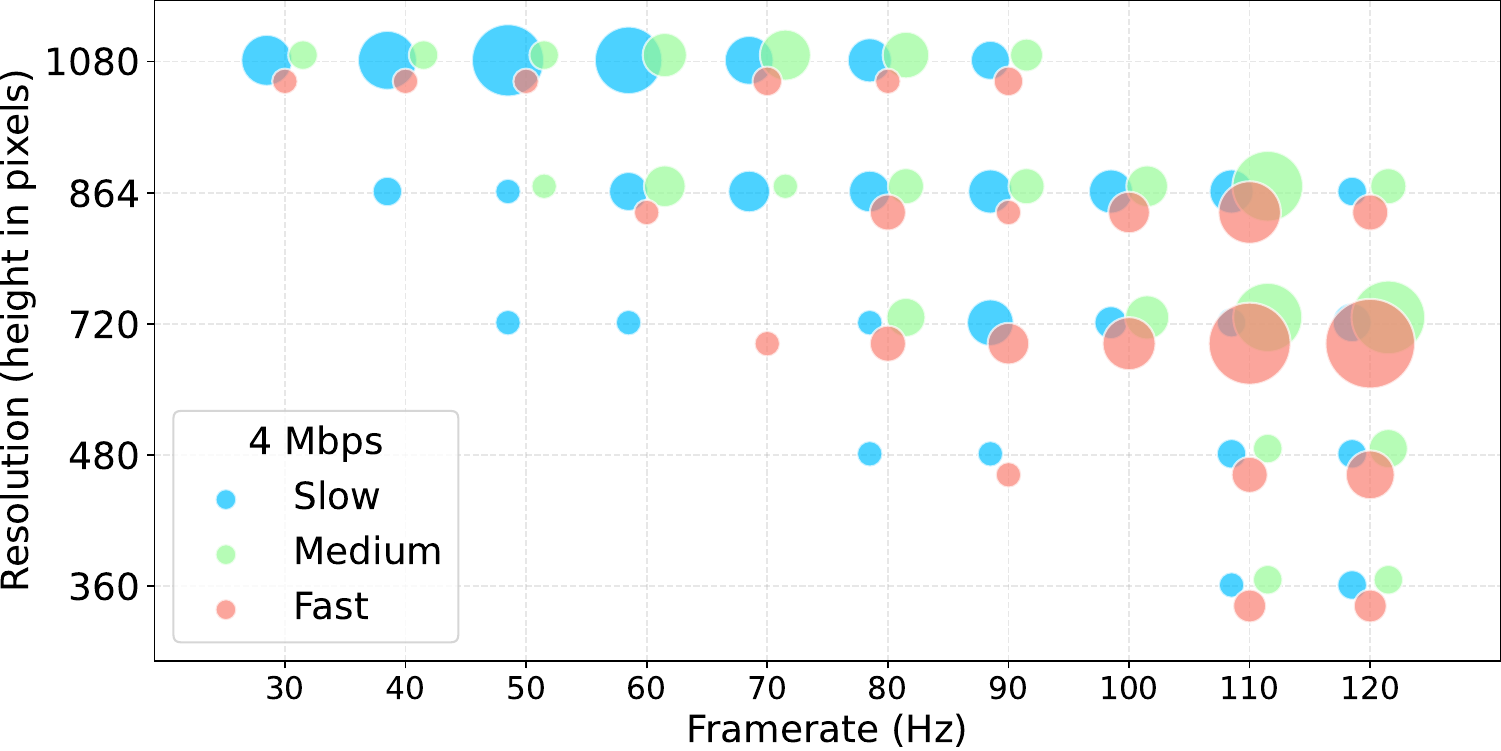}
  \caption{Distribution of (frame rate, resolution) pairs that balance quality and efficiency, achieving quality within 0.25\, JOD units of the optimal (see \eqref{optimal-f-r}) under 2, 3, and 4\,Mbps. The position of each point is offset to better visualize density. Slower motion (blue markers) dominates at lower frame rates. The overall results suggest that a higher frame rate is not always the optimal choice.}
\label{fig:velocity_scatter}
\end{figure}

\figref{velocity_scatter} shows the distribution of selected refresh rates and resolutions across all clips, plotted separately for three bit rates. We can observe an expected bias towards higher frame rates and resolutions, with a higher bias at higher bandwidth. The lower refresh rates and resolutions are associated with lower velocities. 

\subsection{Frame rate and resolution predictor}
\label{sec:predictor}
Because of the complex input (rendered patches, velocity, bandwidth) and simple output (frame rate and resolution), the natural choice for a predictor is a neural network with a combination of convolutional and fully connected layers. Our frame rate and resolution predictor (FRRP) is a compact model with only 293,103 parameters, capable of real-time inference (2\,ms). 
We do not need to query the predictor every frame (see \secref{viterbi}); therefore, inference can run in a separate thread across multiple frames, which makes it suitable for integration into real-time rendering applications.

The architecture of the FRRP is shown in the bottom part of \figref{process}. It consists of a patch encoder, responsible for transforming 128$\times$128 patches into 32-value latent code vectors. It employs a 6-layer convolutional neural network (organized into three blocks of two layers each), followed by global average pooling and a 2-layer MLP with ReLU activation. A sigmoid activation in the last layer ensures that latent values are within the [0, 1] range. The latent code is concatenated with the velocity and available bandwidth and then passed to a classifier, which predicts 5 discrete resolutions and 10 discrete refresh rates, corresponding to those used in the dataset.

To compute the velocity, $v$, for the predictor, we first average motion magnitudes across the frame, then we use a moving average (last 500\,ms) to obtain a representative value. The motion vectors are first extracted from the G-buffer in Normalized Device Coordinates (NDC), then converted to degrees per second ($^\circ$/s) to make input to our predictor independent of the physical display. Because the motion distribution is heavily skewed toward low velocities, we apply a log transform to the motion input. This effectively expands low-magnitude motion while compressing high-velocity outliers. Moreover, we use 80$^\circ$/s as the upper bound for motion, as the smooth pursuit eye motion (SPEM) of the human vision is unable to track higher velocities \cite{robinson1964mechanics}.

\begin{figure}
    \centering
    \begin{subfigure}{1\textwidth} 
        \includegraphics[width=0.5\linewidth]{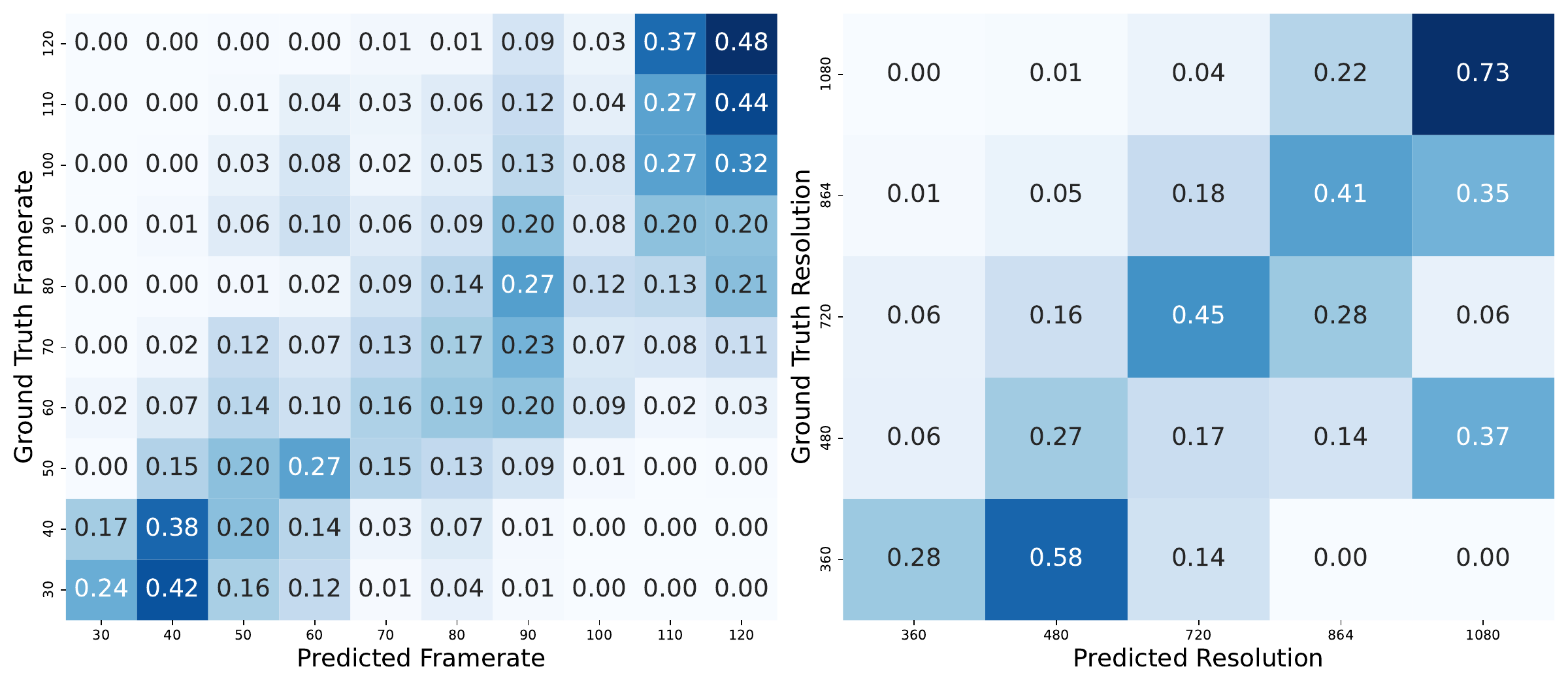}
    \end{subfigure} \\ \vspace{5pt} 
    
    \begin{subfigure}{1\textwidth}
        \includegraphics[width=0.5\linewidth]{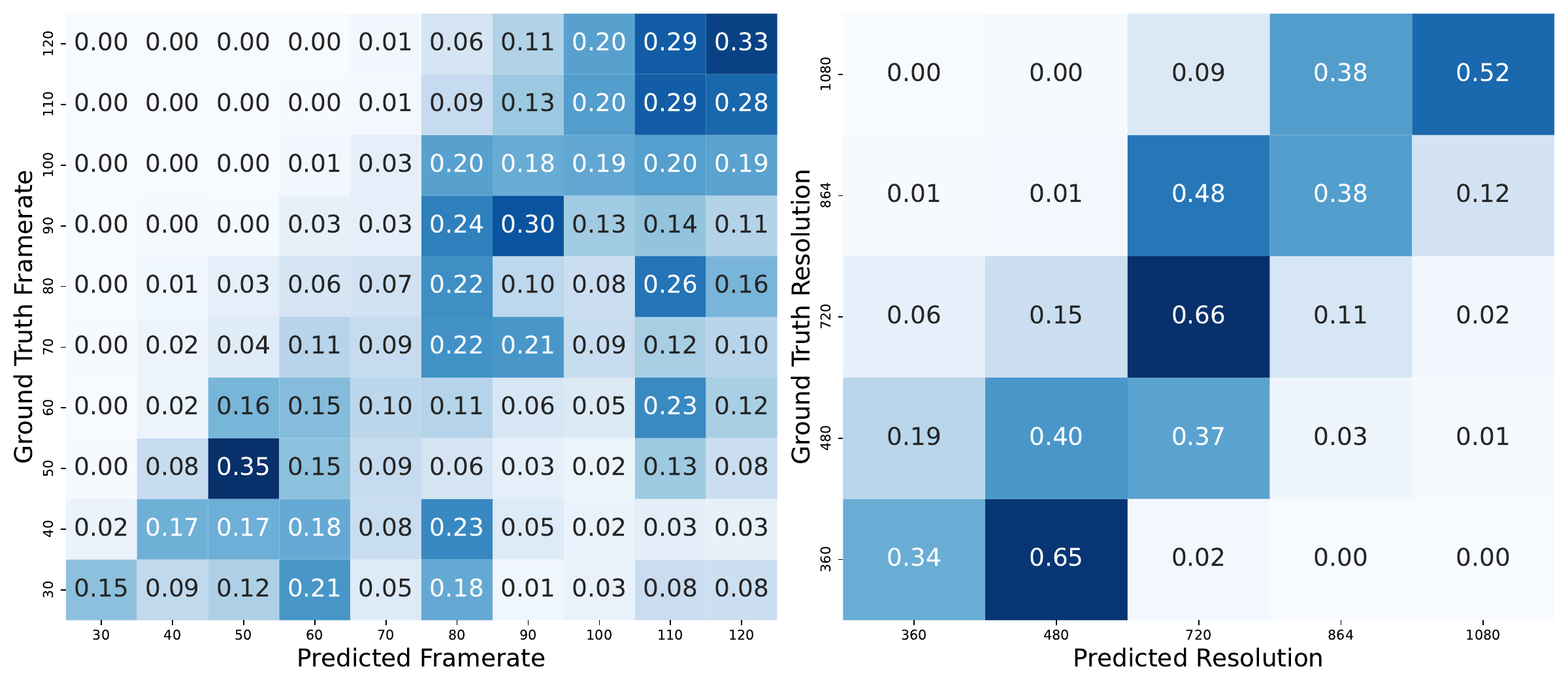}
    \end{subfigure} \\ \vspace{5pt}
    
    \begin{subfigure}{1\textwidth}
        \includegraphics[width=0.5\linewidth]{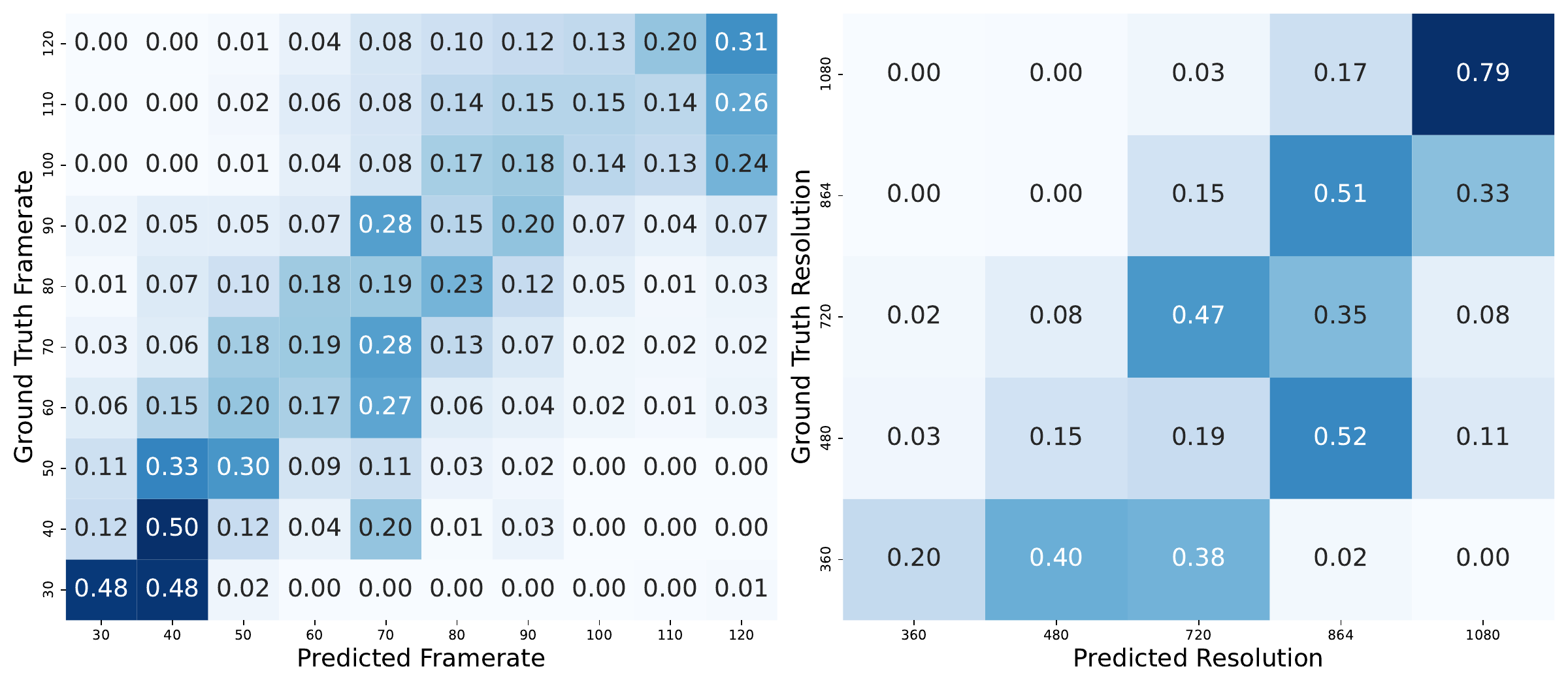}
    \end{subfigure}

    \caption{Confusion matrices for frame rate and resolution prediction, showcasing our 3-fold cross-validation. These results correspond to the best-performing model configuration ("Full model" in Table~\ref{tab:ablation}). Each row is normalized by the total number of samples per class.}
    \label{fig:confusionmatrix}
\end{figure}

\subsubsection*{Training}
Seven rendered and two recorded scenes were used for training, while three rendered and one recorded scene were reserved for testing (see \figref{scenes}). We employed a 3-fold cross-validation strategy during the training to ensure model robustness. The training and testing sets comprised approximately 950,000 and 400,000 patches, respectively. We optimized a cross-entropy loss function using the Adam optimizer with a learning rate of $3\times10^{-3}$. Training was conducted for approximately 65 epochs with early stopping, requiring roughly 8 hours on two Nvidia RTX 4090 GPUs. The resulting confusion matrices are provided in \figref{confusionmatrix}.

\begin{figure}
   \includegraphics[width=0.7\columnwidth]{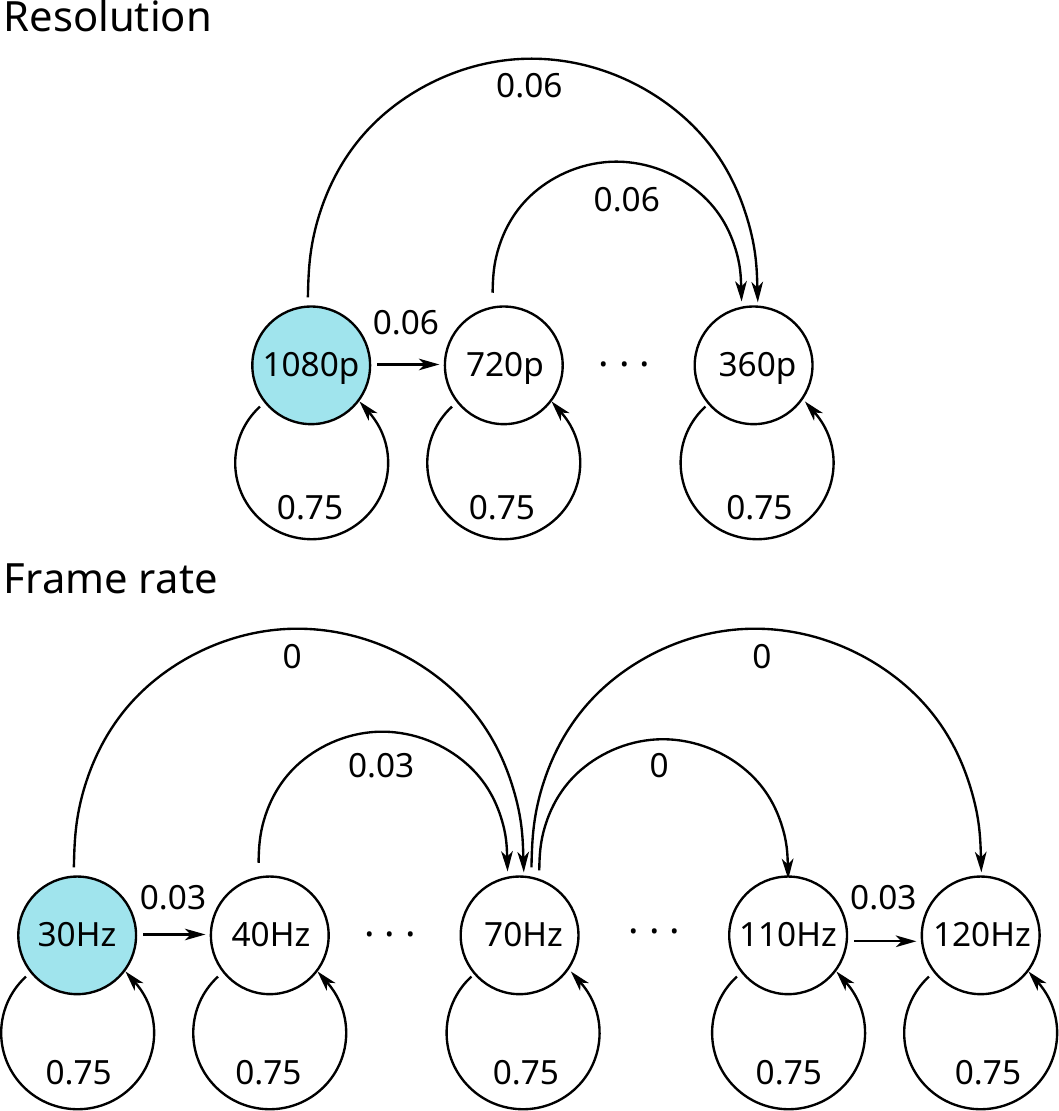}
   \label{Viterbi}
\caption{Transition graphs showing the weights used for the Viterbi algorithm used to control transitions in resolution (top) and frame rate (bottom). The current state of each graph is shown in blue (1080p, 30\,Hz). Note that frame rates which deviate by more than 30\,Hz from the current state have a transition weight of 0, preventing any such transitions.}
\label{fig:viterbi}
\end{figure}

\subsection{Dynamic frame rate and resolution selection}
\label{sec:viterbi}

We cannot switch the frame rate and resolution too frequently for two reasons. First, most video codecs must restart the stream when the resolution is changed, which introduces additional overhead into the available bandwidth. Second, frequent changes in frame rate and resolution can be noticeable, distracting, and affect visual quality. For that reason, we use the Viterbi algorithm to ensure the best frame rate and resolution is selected every 2 seconds. We initialize the Viterbi algorithm with the transition graph weights shown in \figref{viterbi}. The weights are selected to prevent too frequent switching of the refresh rate and resolution. The predictor is run and Viterbi state is updated every frame, but the rendering frame rate and resolution are updated only every 2\,seconds. A typical group of pictures (GOP) — the maximum sequence length between two I-frames — is 1 to 5 seconds. As every resolution update requires inserting an I-frame, it is desirable to align the resolution changes with the start of the GOP, which also introduces an I-frame. In our implementation, we set the HEVC group-of-pictures to 2\,seconds to coincide with potential resolution changes. 

\section{Validation}
\label{sec:subjective_experiment}
We validated our approach through two user studies. The first experiment demonstrates that a JOD reduction of 0.25 from the optimal configuration \eqref{optimal-f-r} does not show a noticeable degradation in perceived quality. The second experiment evaluates our technique against baseline approaches of streaming interactive rendered content. 

\subsubsection*{Display and viewing conditions}
The animations were shown on an LG OLED Evo G2 55-inch 4K display with firmware version 3.0.10. 
The display was connected to a workstation over a 4K HDMI 2.1 cable, which was able to transmit video content at resolutions and frame rates exceeding those used in this experiment. 
A  Windows 11 workstation driving the display was equipped with an Nvidia GeForce RTX 4080 GPU. We conducted the experiment in a dark room. The observer viewed the content at a distance of 190\,cm, corresponding to 39 pixels per degree for 1080p content. 

\subsubsection*{Participants} 10 observers (ages 22-32, 8 males, 2 females) participated Experiment~1. For Experiment~2, ten observers (ages 22–34; 7 males, 3 females) were recruited, including six returning participants. All observers reported that they had normal color vision and either normal or corrected-to-normal visual acuity, with some prior experience with interactive content, such as playing video games.
Both studies were approved by the departmental ethics board, and participants were compensated for their time.

\begin{figure*}[t]
  \centering
  \begin{subfigure}[b]{0.49\linewidth}
    \centering
    \includegraphics[width=\linewidth]{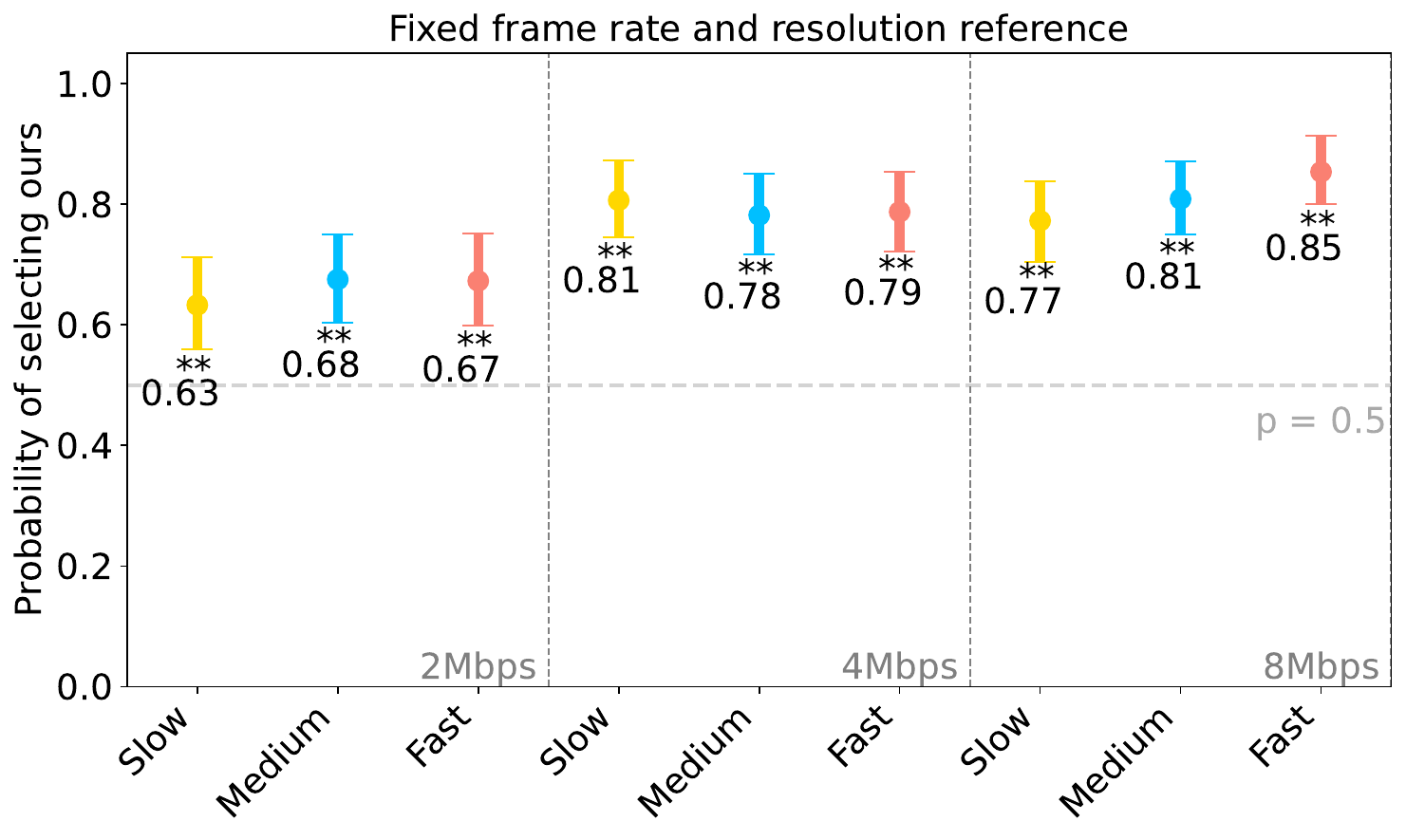}
    \caption{Fixed Reference}
    \label{fig:res_fixed}
  \end{subfigure}
  \hfill 
  \begin{subfigure}[b]{0.49\linewidth}
    \centering
    \includegraphics[width=\linewidth]{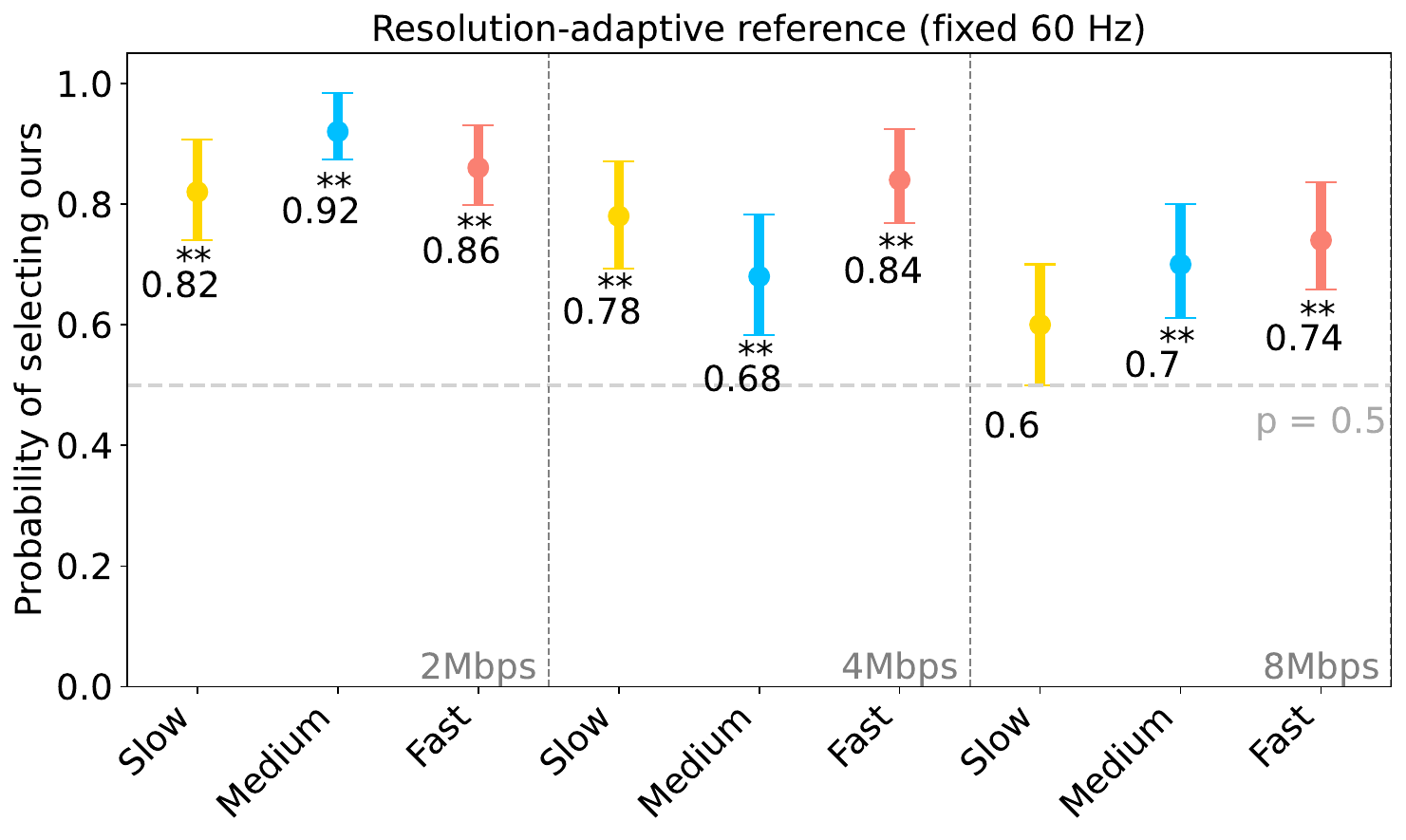}
    \caption{Resolution-Adaptive Reference}
    \label{fig:res_adaptive}
  \end{subfigure}

  \caption{Results of the validation experiment for the two \textit{baseline} conditions: Left --- results against a fixed frame rate and resolution reference. Right --- comparison against a resolution-adaptive reference. The y-axis indicates the proportion of trials in which our method was preferred over the reference videos. Error bars denote 95\% confidence intervals. Results on the x-axis are grouped by motion velocity and bitrate. Asterisks mark statistical significance from a one-tailed binomial test against chance (50\%): * for $p < 0.05$, ** for $p < 0.01$.}
  \label{fig:exp-result-mixed}
\end{figure*}

\subsection{Experiment 1: JOD reduction}
\label{sec:exp-jod-reduction}
We first validate that a 0.25\,JOD reduction in quality is barely perceptible, confirming this threshold does not significantly degrade perceived quality.

\subsubsection*{Stimuli} Animation sequences (3–5\,s) were rendered in real-time using a modified Falcor engine following our dataset generation procedure (\secref{dataset}), with predefeined camera paths to ensure observer evaluated the same content.
Reference videos utilized the maximum ColorVideoVDP quality configuration (round markers in \figref{cvvdp_example}), while test videos were rendered within the 0.25\,JOD margin (triangle markers). To maximize the visible differences, we selected 10 sequences exhibiting the largest $f\,r^2$ discrepancies (see \eqref{optimal-f-r}) between test and reference pairs. 
 
\subsubsection*{Experimental procedure} Observers were shown sequentially pairs of test and reference videos featuring the same content and camera path in a randomized order. They were asked to pick the video with superior perceived quality --- appeared sharper, had fewer distortions, and looked smoother. Both videos were shown on the same display, and observers could freely switch between them without time constraints; each switch triggered a 500\,ms blank and a reset of the camera sequence and encoder state to ensure a fair comparison. Each observer completed a total of 10 comparisons.

\subsubsection*{Results}
Across all trials and all observers, the probability that the reference (maximum quality) is selected over the test (reduced JOD) was 0.55. Given the null hypothesis of random guess ($H_0:\pi=0.5$), we tested for an alternative hypothesis that the reference videos were preferred ($H_\ind{A}:\pi>0.5$). The binomial test (one-tailed, $N=100$, $\alpha=0.05$, $p=0.136$) did not provide evidence for rejecting $H_0$, therefore, we have no evidence indicating that the videos with the JOD difference of less than 0.25 result in lower video quality. The test does not protect from Type-II errors. However, given the small effect size, achieving a statistical power of 0.8 would require more than 770 comparisons, which is impractical to collect.


\subsection{Experiment 2: System validation}
\label{sec:experiment2}
Second, we validate whether our technique improves the quality against constant frame rate. 
The experiment follows the same procedure as Experiment~1, with the following differences:

\subsubsection*{Stimuli} We tested our techniques on a diverse set of scenes, each 8--10\,s long, that included static and dynamic content, varying scene complexities and motion velocities. We used all scenes from the test portion of our dataset (see \figref{scenes}), except for the ``Living room'', which lacked texture complexity. Scenes ``Sibenik'', ``Make Way'', and ``Rogue Company'' have complex geometry, ``Breakfast room'', ``Salle de bain'', and ``Pro Skater'' feature flat-shaded surfaces, which are prone to banding artifacts after video encoding. ``School'', ``Make way'', and ``Rogue Company'' contain dynamic objects and cameras, while the others contain only camera animation. To ensure consistency, all sequences used predefined camera paths. Motion magnitude was extracted from the G-buffer for Falcor-rendered scenes and estimated via Farnebäck’s dense optical flow \cite{farneback_two-frame_2003} for recorded content.

We compared our technique against baselines with fixed resolution and refresh rate, and against a baseline with adaptive resolution but fixed refresh rate.
The first baselines are based on the recommendations set out in \cite{nvidia_obs_nodate}: $1280\times720$ at 60\,Hz for bitrates below 5\,Mbps, and $1920\times1080$ at 60\,Hz  for bitrates above 5\,Mbps.
We chose the above configurations because these two resolutions are most commonly used for game streaming. Although $1920\times1080$, at 120\,Hz can provide the best quality when streamed at high bitrates, it will also put an excessive load on the GPU, and it brings little improvement in quality at lower bitrates due to coding artifacts. Since we target``standard tier'' systems, for which GPU utilization and bandwidth are a concern, 120\,Hz is not a suitable baseline for us.
The comparisons with those baselines included 3 scenes × 3 velocities × 3 camera paths × 3 bitrates = 81 pairwise comparisons using rendered scenes per observer. An additional 24 comparisons were collected using recorded gaming content. The second baseline was a resolution-only adaptive predictor (fixed at 60\,Hz) that we trained separately. The purpose is to see if adaptively selecting both frame rate and resolution yields better results than adapting resolution alone. The comparison with this baseline included 3 scenes × 3 camera paths with different velocities × 3 bitrates = 27 comparisons for rendered content, plus 24 comparisons for recorded content.

\subsubsection*{Results}
Results in Fig.~\ref{fig:exp-result-mixed} show that our approach improves perceptual quality across nearly all bitrates, with the clearest improvement above 2\,Mbps compared to the fixed-baseline, and below 4\,Mbps compared to the resolution-adaptive baseline. We speculate that this is due to the lower bit rate introducing a larger level of overall distortion into all video sequences thus making quality judgments somewhat harder to form. 

\section{Ablations}
\label{sec:ablations}
We evaluate the predictor's design through ablation studies, with performance results reported in Table~\ref{tab:ablation}. The reported errors are averaged over three train/test splits (folds), where each split included one screen-recorded and three scenes from Falcor. The remaining scenes were used for training. Those splits are separate from the train/test split used in \secref{subjective_experiment}.
Table~\ref{tab:ablation} shows that removing either velocity or patch information decreases performance. 
Velocity has a bigger impact on the frame rate prediction, while the input patch more strongly influences resolution prediction --- an expected result given the spatial nature of resolution.
Increasing the patch size from 32 to 128 significantly reduces prediction error, as larger spatial areas provide the network with more context to distinguish between high-frequency textures and aliasing.

The relative error metric used in \tableref{ablation} is given by the equation:
\begin{equation}
E = \left( \exp \left( \frac{1}{n} \sum_{i=1}^{n} \left| \log(R_{\text{test},i}) - \log(R_{\text{gt},i}) \right| \right) - 1 \right) \times 100 
\label{eq:relative_error}
\end{equation}

The confusion matrices for the best-performing model are shown in \figref{confusionmatrix}. They show that in the majority of cases, the prediction error is just one class --- one step in the resolution or refresh rate. Such misprediction is unlikely to cause noticeable degradation of performance.

\begin{table}[t]
\caption{Ablation study on input configurations. We evaluate the impact of patch size, number of patches, and velocity information on model performance. The Full model (bolded) achieves the best overall performance across all error metrics with 293,103 parameters, making it suitable for real-time applications (< 2\,ms). 
}
\renewcommand{\arraystretch}{0.98} 
\scalebox{0.75}{
\begin{tabular}{ p{0.26\columnwidth}  p{0.2\columnwidth}  p{0.16\columnwidth}  p{0.16\columnwidth} p{0.14\columnwidth}  p{0.16\columnwidth} }
  \toprule
  \textbf{Model} & \textbf{Patch size}  & \textbf{$n_{patches}$}& \textbf{Velocity} & \textbf{FPS error} & \textbf{Resolution error} \\ 
  \midrule
  w/o patch & NA & NA & \checkmark & 22\% & 19\% \\ 
  w/o velocity & 128 & 1 & \xmark & 33\% & 24\% \\ 
  1$\times$32$\times$32 & 32 & 1 & \checkmark &  45\% & 21\% \\ 
  1$\times$64$\times$64 & 64 & 1 & \checkmark & 41\% & 18\% \\ 
  \textbf{Full model} & \textbf{128} & \textbf{1} & \checkmark & \textbf{22\%} & \textbf{17\%} \\
 \bottomrule
\end{tabular}}
\label{tab:ablation}
\end{table}

In addition to the above ablations, we experimented with a predictor that processed multiple randomly selected patches from a frame and combined their information by average-pooling their latent vectors $\mathbf{e}$. We also tried replacing a classifier with a regressor. In both cases, the predictor's performance was comparable or worse. 

\section{Conclusions}
Higher frame rates significantly enhance computer graphics, particularly in high-motion content. This is because they reduce latency \cite{Spjut_2019}, mitigate blur and judder \cite{Watson_2013,denes_perceptual_2020}, and generally improve visual quality. Our work shows that real-time game streaming services can exploit these benefits by adaptively selecting both the frame rate and the resolution. 

Optimizing these parameters in real-time is challenging due to the complex interplay among factors like textures, velocity, compression artifacts, and spatio-temporal sensitivity. To address this, we successfully adapted a video metric for this task, enabling the labelling of a large dataset of 69,611 video clips, and training a neural network that predicts the optimal rendering parameters in real-time. While our primary goal was to maximize visual quality under bandwidth constraints, permitting a slight reduction in predicted quality yields substantial reductions in rendering cost.
By leveraging the spatiotemporal limits of human vision, these computational savings are achieved while preserving high perceived quality.

\subsubsection*{Limitations} 
Modern games often mix dynamic rendered content with static user interface overlays, such as a mini-map or status information. Our method optimizes for the dynamic content and may lower the quality of user interface elements. Ideally, those should be rendered and streamed separately from the dynamic content, or our method should control shading rate rather than resolution.

\edit{Reducing GPU utilization in game streaming is often highly desirable for service providers as it can reduce overall power consumption or allow more virtual GPU (vGPU) instances to run on a single GPU. Our method also improves video quality in bandwidth-constrained scenarios. However, our solution does not address the issues caused by network latency, which can have a substantial impact on the quality of experience.} 

\edit{Our dataset contains mostly indoor environments without transparency, particle effects, dynamic lighting changes or stylized/non-photorealistic shading. Such content may degrade the performance of our predictor in cases where on-screen motion is not captured by motion vectors stored in the G-buffer (dynamic lighting changes, transparent particle systems, etc.).}

While our modification of ColorVideoVDP proved effective at the task, as shown in two validation experiments, the metric can be further refined by retraining on a suitable dataset.

\begin{acks}

We would like to thank anonymous reviewers for their valuable feedback on our work. We also thank experiment participants for their contribution to this research. 

\end{acks}


\bibliographystyle{ACM-Reference-Format}
\bibliography{sample-base}

\clearpage
\newpage

\end{document}